%% file: ms.tex
\documentclass[12pt,preprint]{aastex}

\received{}
\begin{document}

\title{Optimal Image Reconstruction in Radio Interferometry}

\author{E. C. Sutton}
\affil{Department of Astronomy, University of Illinois, 1002 W. Green St., Urbana, IL 61801}
\author{B. D. Wandelt}
\affil{Departments of Physics and Astronomy, University of Illinois, 1110 W. Green St., Urbana, IL 61801}
\email{sutton@astro.uiuc.edu, bwandelt@uiuc.edu}

\begin{abstract}

We introduce a method for analyzing radio interferometry data
which produces maps which are optimal in the Bayesian sense of
maximum posterior probability density, given certain prior assumptions.
It is similar to maximum entropy techniques,
but with an exact accounting of the multiplicity instead of
the usual approximation involving Stirling's formula.
It also incorporates an Occam factor, automatically limiting the effective
amount of detail in the map to that justified by the data.
We use Gibbs sampling to determine, to any desired degree
of accuracy, the multi-dimensional posterior density distribution.
From this we can construct a mean posterior map and
other measures of the posterior density, including confidence limits
on any well-defined function of the posterior map.

\end{abstract}


\section{INTRODUCTION}

A radio interferometer samples the Fourier transform
of the sky brightness distribution.
This statement oversimplifies the situation \citep{TMS}, 
but is sufficient to illustrate the problem of reconstructing images
of the radio sky.
The brightness distribution $I(\xi,\eta)$ is the inverse Fourier transform
of the visibility function $V(u,v)$,
\[
I(\xi,\eta) \rightleftharpoons V(u,v) .
\]
Here $\xi$ and $\eta$ are directional cosines
in the directions of increasing right ascension and declination,
respectively, with respect to some phase center.
The spatial frequency coordinates $u$ and $v$ are components of the projected
baseline divided by the wavelength, conjugate to $\xi$ and $\eta$, respectively.
Since $V(u,v)$ is of necessity incompletely sampled, 
the intensity distribution $I(\xi,\eta)$ cannot be recovered uniquely
by the inverse Fourier transform of $V(u,v)$.
Instead, one initially can recover what is known as the principal solution,
the inverse Fourier transform of the visibility times the sampling function.
This results in a brightness distribution (the dirty map)
consisting of $I(\xi,\eta)$ convolved by the inverse Fourier transform
of the sampling function (the dirty beam).
Depending on the $u-v$ sampling,
the dirty beam will contain artifacts
such as grating rings and sidelobes, which can be troublesome.
The problem then is to recover a better view of the radio sky than
this principal solution.  We use the
term principal solution whenever the visibility is set to zero
at unsampled points, regardless of any weighting applied to
the sampled data.

The procedure CLEAN, used to address this problem, was
first described by \cite{H74}.
The technique may be visualized by considering
the sky as containing a finite assembly of point sources.
The algorithm sequentially identifies the strongest source,
removes the beam pattern produced by such a source,
and iterates until the noise level is reached. 
The implicit prior 
is that the sky brightness distribution is zero over most of the image.
Subsequently, attention was given to
techniques variously known as phase closure, self calibration,
and hybrid mapping \citep{RW78,CW81,SC83},
techniques which were effective at removing phase
and amplitude instabilities in interferometry data.
It is usually assumed that the complex gain for a given baseline
can be factored into complex antenna-based gains,
\[
\tilde{G}_{\alpha\beta} = \tilde{g}_{\alpha} \tilde{g}_{\beta}^{\ast} \; .
\]
The problem is more complex if the passbands differ enough
so that such factorization is not possible.
CLEAN, as described above, exhibits a striping instability
in regions of extended emission.
\cite{C83} and \cite{S84} addressed this problem,
although in different fashions.
The situation through 1984 was summarized by \cite{PR84}.
\cite{BS84} and \cite{WS88} subsequently introduced
a multi-scale approach to CLEAN.

Other work concentrated on the problem of the
missing "zero-spacing" data, which produces bowls
of negative emission (depressions) around strong sources.
Since an interferometer cannot observe with zero baseline,
the principal solution map always contains zero net flux.
A solution is to augment interferometric data with
either total-power or mapping data from large single-dish telescopes.
This problem was addressed by \cite{BvA79,ER79,V84,M88,C88}.
This topic encroaches upon other deconvolution methods,
such as maximum entropy, and is discussed below.
It also is related to mosaic mapping \citep{S96}, also discussed below.

CLEAN has become the best established method
of reducing radio interferometer data,
despite its proclivity to reproduce point sources well
while poorly reproducing regions of extended emission.
We believe this is mostly a matter of habit among astronomers,
who have become attached to the intuitive nature of CLEAN.
However, it contains some troubling features
including its non-linear nature, the implicit emptiness prior,
and a number of \emph{ad hoc} choices 
such as the loop gain, the restoring beam, and the stopping criteria.
A non-linearity cannot be avoided if an improved map is to be obtained.
The first and the third of these ad hockeries
(to use the splendid coinage of \cite{J82,J03}) 
seem to have little practical consequence,
although the \emph{ad hoc} restoring beam sometimes prevents observers
from seeing the full resolution (superresolution) justified by
high signal-to-noise data.

Even in the original work of \cite{H74}, the main advantage of CLEAN was
described as the ease of computation.  H\"{o}gbom seemed to anticipate
the advantages of maximum entropy algorithms, even using language
remarkably Bayesian in character.  Computing power may have been a
sufficient justification in favor of CLEAN in 1984. 
But lack thereof is certainly 
insufficient justification for current and future generations
of radio interferometers such as CARMA and ALMA.

The maximum entropy method (MEM) introduced 
by \cite{A74} was developed for image restoration
by \cite{GD78,BS80,GS84,CE85} as an alternative to CLEAN.
The technique also has found great success in 
crystallography, geophysics, tomography,
and a variety of other applications.
A review of MEM in the context of image restoration
in astronomy was given by \cite{NN86}.
The basic idea of MEM is to provide an image which has the
smallest amount of structure, the maximum entropy, while remaining
consistent with the observed visibilities within observational uncertainties.
Much of this work used the configurational entropy
as the appropriate measure of entropy.
We write this as
\[
S = - \sum_{i} \frac{I_{i}}{I_t} \: ln \, \frac{I_{i}}{I_t} \; ,
\]
where the summation is over all pixels and
\[
I_t = \sum_{i} I_{i} .
\]
The normalization with $I_t$ keeps
the argument of the logarithm dimensionless.
In this formulation the entropy has natural units,
known variously as "nits", "nats", or "nepers"
(the units would be "bits" if $log_2$ had been used).
There is no multiplicative constant,
as this would simply correspond to a change of units
and a change in the base for logarithms and exponentiation.
This is important in our analysis.
We do not include the concept of default image models,
since we believe they would be difficult to specify reliably
in astronomical imaging.

The penalty function is taken to be
\[
\frac{\chi^2}{2} - \lambda S
\]
where $\lambda$ is a Lagrange multiplier, and the penalty
function is to be minimized, subject to some constraint.
The term $\lambda$ is often viewed as a control parameter,
adjusting the degree of smoothness imposed on the solution.
From an information-theoretic point of view and
following \cite{J82}, we show below that this constrained
optimization problem may be viewed equivalently as an unconstrained
Bayesian maximization.  This assumes that the desired result
is the map with maximum \emph{a posteriori} probability density.
We also show that it is possible to obtain important
additional information about the posterior probability density distribution.
The value of $\lambda$ should be determined by the data.
However, if the magnitude of the errors in the experimental
data are not know well, then the scaling of $\chi^2$ is unknown
and a Lagrange multiplier will be needed.
The total flux in the map is $I_t$ times the solid angle per pixel.
Some approaches include an additional Lagrange multiplier for the total flux.
If the total flux is known from one or more single-dish measurements,
they may be included as terms in $\chi^2$.
If not, it is hard to see how to justify such a term.

\cite{M88} and \cite{C88} extended the MEM approach to
include joint deconvolution of single-dish and interferometer data.
Our work here is a continuation along this general line,
merging such considerations with the maximum entropy approach
of \cite{GD78} into a proper Bayesian framework.
It also allows for a simple inclusion of mosaic data.

\section{BAYESIAN ANALYSIS}

From the beginning of this development, terms like "prior" and
"a priori information" have appeared frequently in the literature
without, however, many attempts at formal Bayesian analysis.  
The MEM discussions of \cite{GD78} and \cite{J82}
have come closest.
Bayes theorem may be derived from the product rule,
\[
P(AB \vert C) = P(A \vert BC) \; P(B \vert C)
\]
\[
P(AB \vert C) = P(B \vert AC) \; P(A \vert C) .
\]
$P(AB \vert C)$ stands for the conditional probability of A and B,
given C, etc.
Combining these two equations we get Bayes equation,
\[
P(H \vert DI) = P(H \vert I) \; \frac{L(D \vert HI)}{P(D \vert I)} \; ,
\]
where we substitute H (standing for hypothesis) for A,
D (standing for data) for B, and
I (standing for information) for C.
The information I represents our combined knowledge about 
such things as beam patterns, measurement errors, etc.
Our various hypotheses are maps, and we wish
to find maps with high probability.

The left-hand side of the equation is referred to as the posterior.
The first term on the right-hand side is referred to as the prior.
The numerator of the second term is the likelihood,
indicated by the symbol L, and the denominator (the "evidence")
is a normalization factor which sometimes may be ignored.
The interpretation of Bayes theorem is that the posterior
probability of any hypothesis H is given by the product of one's
prior knowledge of that hypothesis times the likelihood that the
hypothesis would have produced the observed data (suitably normalized).

\subsection{Likelihoods}

The likelihood of the data, given the model and the information,
can be factored into multiple terms if the data are independent.
This is nothing more than
what one does in calculating $\chi ^2$, where each datum is
represented by a multiplicative term in the likelihood and therefore
an additive term in the logarithm of the likelihood ($\chi ^2 /2$).
Here we factor the likelihood into two terms:
$L_{int}$ for the observed set of interferometric data (visibilities)
and $L_{sd}$ for any image-domain measurements
obtained from single-dish pointings or mappings.
\[
L ( \{Data\} \vert I(\xi,\eta) \, I ) =
L_{sd} ( \{T_k\} \vert I(\xi,\eta) \, I ) \times
L_{int} ( \{V_k\} \vert I(\xi,\eta) \, I ) \; .
\]
The single-dish likelihood is the product, over various pointings
of the telescope towards celestial offset coordinates $\{\xi_k, \eta_k\}$,
of the likelihood of each measured antenna temperature in $\{T_k\}$.
For simplicity we assume the single dish beam pattern to be
symmetrical and Gaussian (a property we also assume for
the individual telescope beams in the interferometers).
Generalization to more complex patterns is straightforward \citep{AW04}.
We describe the antenna pattern in terms of the
effective collecting area, $A_\mathit{eff}$,
as a function of angles with respect to the telescope pointing direction.
For a given map $I(\xi,\eta)$, the expected antenna temperature is
\[
T_k^{\prime}(\xi_k,\eta_k) = \frac{1}{2k_b} \; \int\!\int I(\xi,\eta) \;
A_\mathit{eff}(\xi-\xi_k,\eta-\eta_k) \, d\xi \, d\eta \; .
\]
The symbol $k_b$ represents Boltzmann's constant.
We assume additive, independent Gaussian noise with probability distribution
\[
P(N) = \frac{1}{\sqrt{2\pi}\sigma_k}\;e^{-N^2/2\sigma_k^2}
\]
\[
\sigma_k = \frac{K\,T_{sys}}{\sqrt{\Delta\nu\,\tau}} \; ,
\]
where K is a constant of order unity that depends on the method
of observation (e.g. beam switching), $T_{sys}$ is the system
noise temperature, $\Delta\nu$ is the bandwidth, and $\tau$ is
the integration time.  The noise $\sigma_k$ may vary between measurements.
The likelihood of the observed $\{T_k\}$ is
\[
L_{sd} ( \{T_k\} \vert I(\xi,\eta) \, I ) =
\prod_{k=1}^{N_{sd}} \; \frac{1}{\sqrt{2\pi}\sigma_k} \;
e^{-(T_k - T_k^{\prime})^2/2\sigma_k^2}
\]
and the log likelihood (ignoring constant terms) is
\[
ln \; L_{sd} ( \{T_k\} \vert I(\xi,\eta) \, I ) =
-\frac{1}{2} \sum_{k=1}^{N_{sd}} \frac{(T_k - T_k^{\prime})^2}{\sigma_k^2} \; .
\]

The interferometric likelihood is the product, over the various
visibility measurements in $\{\tilde{V}_k\}$, of the likelihood of each such
visibility at spatial frequencies $\{u_k,v_k,w_k\}$, given a map $I(\xi,\eta)$.
We assume the bandwidth pattern to be unity, as it would be for
negligible bandwidth, or to be incorporated into the antenna pattern.
The expected visibility is
\[
\tilde{V}_k^{\prime}(u_k,v_k,w_k) =
\int\!\int P_{N}(\xi,\eta) \, I(\xi,\eta) \,
\frac{exp\{-i2\pi[\xi u_k+\eta v_k+w_k(\sqrt{1-\xi^2-\eta^2}-1)]\}}{\sqrt{1-\xi^2-\eta^2}}
\; d\xi \, d\eta \; ,
\]
where $P_N$ is the normalized power pattern of an individual telescope,
given by
\[
P_{N}(\xi,\eta) = A_\mathit{eff}(\xi,\eta) / A_\mathit{eff}(0,0) \; .
\]
Since the interferometric antennas are not
generally the same as the single dish antennas,
the symbol $A_\mathit{eff}$ can take on different meanings.
We believe the meaning will be clear, depending on the context.
If the telescopes of an interferometric pair have different power patterns,
the geometric mean of their power patterns should be used,
\[
P_{N} = \sqrt{P_{\alpha}P_{\beta}} \, .
\]
Note the use of the exact kernel in the transform integral,
rather than the Fourier kernel, which is only approximate.
Here we assume additive, independent, bivariate Gaussian noise
with distribution
\[
P(\tilde{N}) = \frac{1}{2\pi\sigma_k^2} \; e^{-(Re\tilde{N})^2/2\sigma_k^2} \; e^{-(Im\tilde{N})^2/2\sigma_k^2}
\]
where
\[
\sigma_k = \frac{2k_bT_{sys}}{A_\mathit{eff}\,\eta_Q} \; \frac{1}{\sqrt{\Delta\nu\,\tau}} \; .
\]
The quantization efficiency factor $\eta_Q = 0.881$
for 2 bit (4 level) quantization at the Nyquist rate \citep{TMS}.
The system temperature is the geometric mean of the system
temperatures of the pair of telescopes involved in the observation,
\[
T_{sys} = \sqrt{T_{\alpha}T_{\beta}} \, .
\]
The likelihood of the observed $\{\tilde{V}_k\}$ is
\[
L_{int} ( \{\tilde{V}_k\} \vert I(\xi,\eta) \, I ) =
\prod_{k=1}^{N_{int}} \frac{1}{2\pi\sigma_k^2} \;
e^{-|\tilde{V}_k - \tilde{V}_k^{\prime}|^2/2\sigma_k^2}
\]
and the log likelihood (ignoring constant terms) is
\[
ln \; L_{int} ( \{\tilde{V}_k\} \vert I(\xi,\eta) \, I ) = - \frac{1}{2}
\sum_{k=1}^{N_{int}} \frac{|\tilde{V}_k - \tilde{V}_k^{\prime}|^2}{\sigma_k^2} \; .
\]
For mosaic mapping, all that is necessary is to incorporate pointing
offsets into the definition of $\tilde{V}_k^{\prime}$, just as was
done for $\tilde{T}_k^{\prime}$.

\subsection{Matrix Formulation}

For many purposes the most convenient method of calculating likelihoods
is to formulate the problem as a matrix multiplication
\[
D = M I + \epsilon .
\]
The image $I$ of the sky is represented by a vector enumerating
the values of the various pixels, with some specified ordering of
the $n$ pixels.  The measurement matrix $M$ converts
the image into a variety of measurable quantities $D$ (data), which
may include both visibility data and single-dish data.
Noise $\epsilon$, assumed to be Gaussian with zero mean,
is added in the measurement process.  In the absence of other information
about the sources of noise, the choice of a Gaussian is least informative.
We do not attempt to invert this equation.  Rather, if the measurement
matrix is known exactly and the statistical properties of the noise
are known exactly, one can use this directly to calculate the log likelihood
\[
ln \; L = -\frac{1}{2} \sum_{k=1}^{N_{data}}
\frac{| D_k - M_{kj} I_j | ^2}{\sigma_k ^2} .
\]
This formulation is general enough to include single-dish data,
interferometric data, and mosaic data.
The measurement matrix may contain some number of
unknown parameters, M($\alpha , \beta , \gamma , \cdots$),
whose distributions we also seek to discover.

\subsection{Priors}

We seek a pixel-based description of the intensity distribution.
The pixels are square, of equal size, and uniformly distributed.
The region likely to contain relevant information is circular and
approximately the size of the FWHM of the main telescope beam. 
The region of support of the intensity distribution (the region mapped)
should have a diameter approximately twice that of the primary telescope beam.
If there is a strong source well out into
the tails of the primary beam, the support region may be
enlarged as necessary.  For a heterogeneous array, the support should
be twice the geometric mean of the main beams of the two smallest telescopes.
The pixel size should be chosen to allow for some degree
of superresolution, with several pixels across the synthesized beam.

What prior information is available?  The intensity is real,
which means that the visibilities must be Hermitian
(ignoring the $w_k$ terms in the transform kernel). 
This fact is not explicitly required here,
and each observed visibility may be considered only once,
at its conventional $u,v$ location.
The intensity in each pixel is also non-negative.
A positive, additive distribution of this sort
should be treated with an entropy prior \citep{Sivia96}.
This is done below, although in a manner somewhat different than
the conventional approach.

Following \cite{J82} and \cite{GD78}, we consider the intensity distribution
as if it were constructed by the conventional "team of monkeys" throwing
$\lambda$ "elements of luminance" among the various pixels. 
However, unlike some authors we do not take the limit
$\lambda \rightarrow \infty$ nor necessarily the case of $\lambda$
much greater than the number of pixels.  
Therefore we retain the key feature of brightness quantization \citep{J82,GD78}.
The number of such "elements of luminance" and the level of
brightness quantization (q) will be determined by Bayes theorem.
No physical quantization is implied; these are not photons.

For distributing $\lambda$ elements among n pixels,
the multiplicity of some particular distribution \{$N_i$!\} is given by
\[
W = \frac{\lambda!}{N_1!N_2! \cdots N_n!} .
\]
At this point most authors take the limit $ \lambda \rightarrow \infty$
and approximate $\lambda!$ and \emph{all} of the \{$N_i$!\}
using the leading term in Stirling's formula, yielding
\begin{eqnarray*}
ln \, W & = & \frac{1-n}{2} \, ln \, 2\pi + \frac{ln\,\lambda}{2} - \frac{1}{2}\sum_{i}ln \, N_i - \lambda \sum_{i} \frac{I_{i}}{I_t} \, ln \frac{I_{i}}{I_t} \\
 & \approx & \lambda \, S \; ,
\end{eqnarray*}
where S is the logarithmic form
of the configurational entropy discussed in section 1.
We will not use Stirling's formula, since many of the \{$N_i$!\}
may be of order unity.
For this reason we prefer to refer to ours as a \emph{multiplicity} prior,
to differentiate it from the conventional \emph{entropy} prior.
Also note that in the $\lambda \rightarrow \infty$ limit,
the posterior density becomes
proportional to $e^{\lambda S-\chi^2/2}$ \citep{J03}.
This results in a perfectly flat distribution
(the prior overwhelms the likelihood), further illustrating
that the use of Stirling's formula is inappropriate in this context.
Instead, $\lambda$ should be treated as a finite term
whose value is to be determined by Bayes theorem.

Since there are $n^\lambda$ possible configurations,
the properly normalized prior distribution is
\[
p(\{N_i!\}) = \frac{W}{n^\lambda}.
\]
As $\lambda$ gets larger, the prior probability drops due to the
normalization factor of $n^\lambda$ in the denominator.
This is the automatic application of Occam's razor which
occurs in all Bayesian calculations.  Any set of data
allows only a certain amount of detail in the posterior,
corresponding to the information content of the data.

We take a joint prior on the quantization level (q),
the number of luminance elements ($\lambda$),
and their distribution (\{$N_i$\}) and factor it as
\[
p(\{N_i\} \lambda q \vert I) = p(\lambda q \vert I) \; p(\{N_i\} \vert \lambda q I) .
\]
The product $\lambda q$ is a scale parameter
and should be given a Jeffreys prior.
It also represents the total flux in the map.
In our reconstructions the total flux is generally constant,
with $\lambda$ varying approximately as $q^{-1}$.
So for simplicity we consider the term $p(\lambda q \vert I)$
to be constant.  
The use of a Jeffreys prior would make little difference.
The logarithm of the map prior (neglecting constants) is
\[
ln \; p(\{N_i\} \lambda q \vert I) =  ln \, W - \lambda \, ln \, n.
\]
Therefore the logarithm of the posterior is
$ln \, W - \lambda \, ln \, n - \chi^2/2$.
According to Bayesian logic, one then should vary $\lambda$, q,
and the \{$N_i$\} to explore the posterior density.
The posterior of the image is then obtained by marginalizing q and $\lambda$.

For a clear and interesting discussion of the question of
the appropriate $\lambda$, see chapters 5 and 6 of \cite{Sivia96}.  
We believe that our treatment of $\lambda$ meets the requirements
of \cite{NN86}, who described this as an unsolved problem.
Our treatment of $\lambda$ differs from and is preferable to 
that of \cite{GD78}, who chose $\lambda$ so that $\chi^2 = N_{data}$,
a reasonable but \emph{ad hoc} choice.
Finally, take note of the cautions given by \cite {J03},
who warns about the transition from the discrete to
the continuum case and paradoxes that can arise with the latter.

Specification of prior information is the trickiest part
of Bayesian analysis.  Failure of the Bayesian technique is
almost always due to the failure to incorporate all relevant
prior information or to the incorporation of incorrect prior information.
We have spoken of the positive-definite nature of the intensity. 
Here we have to be careful.
\cite{B79} pointed out that
"a lot of entropy lies around the low brightness fringe so the choice
of the boundary [zero] is not a trifle."  
\cite{NN86} commented further on this point.
What is generally referred to as "zero"
in radio interferometric maps is not necessarily the true zero in intensity.
Neither interferometric measurements nor position-switched
single-dish measurements are sensitive to a spatially uniform
component of the emission.
So in continuum maps the true zero of intensity may lie
somewhat below the zero of the map.
In spectroscopic mode the problem is further complicated by
the convention of making maps from "continuum-subtracted" data.
Since spectral lines and continuum often have
different spatial distributions, the issue of the true zero
of the intensity -- which relates to their sum -- intimately links
the continuum and the spectral line maps.
We defer the treatment of spectral line mapping to a later time,
treating here only the continuum case.

\subsection{Implementation}

We first perform a multi-dimensional maximization of the posterior density.
Because of the brightness quantization, this is a problem in
combinatorial optimization.  Such problems are notoriously difficult.
Our parameter space has potentially tens of thousands of dimensions.
This precludes an exhaustive search for the global maximum of
the posterior density.  Instead we may attempt to find one or more good
(high probability) local maxima and a statistically valid sampling of
the posterior probability density.

We make an initial guess for the quantization level, q.
Then we pick an initial intensity distribution. 
This may be as simple as an empty (zero) map.  
But under many circumstances it may be
advantageous to start from a better first approximation,
such as a component map produced by CLEAN or a maximum entropy map.
In such cases the map must then be quantized by taking the pixel values,
dividing by q, the flux per quantum, and rounding to an integer value.

The posterior is then maximized by performing a grid search
using the algorithm described in Appendix A.
The quantization level is then also varied, possibly over
a wide range, to find a region in q over which the posterior is maximized.
The general appearance of the posterior map remains
similar over a large range of brightness quantizations.

This technique is computationally intensive in terms of
both memory requirements and calculational speed.
The work described here has been performed on a
single 1 GHz processor of a workstation-class machine with
1 GBy of memory.  However, this type of problem
may be highly parallelized with both computational
and memory requirements spread over a number of processors.
Savings in storage and computation may be achieved
by making careful use of the Hermitian nature of the transform kernel.
For the case of finite values of $w_k$, the kernel may
be factored into two terms, one of which is Hermitian.
The other is non-Hermitian but symmetric.

\subsection{Joint Densities}

Complete information about the image is contained not
just in the optimum posterior map (the mode of the probability density),
but in the multi-dimensional joint probability density of
the quantization level q, and the occupation numbers of all the pixels.
\cite{HM03} discuss in a variety of astrophysical contexts
the advantages of considering the complete posterior distribution.

We use a variant of Gibbs sampling to numerically
estimate the joint posterior density.
The Gibbs sampler is a particular example of 
the Markov-chain Monte Carlo method.
Its advantage is that it splits a multi-dimensional sampling problem
into a set of lower-dimensional sampling problems.
In our case, the univariate conditional densities for individual pixels
are readily calculable from the value of q and the values of all
the other pixels.  These distributions are discrete, and only
a small number of states have non-negligible probabilities.
From an initial state $\{N_i^0\},q^0$ one obtains
an updated state $\{N_i^1\},q^1$ by a scan through the pixels,
successively updating the $\{N_i\}$ by random draws from
\[
p(N_1^1 \vert N_2^0, N_3^0, ... , N_n^0, q^0)
\]
\[
p(N_2^1 \vert N_1^1, N_3^0, ... , N_n^0, q^0)
\]
\[
...
\]
\[
p(N_n^1 \vert N_1^1, N_2^1, ... , N_{n-1}^1, q^0) .
\]
Next, it is necessary to consider the continuous variable q,
for which the conditional density is not readily available.
This variable is updated with a Metropolis-Hastings step using draws
from a Gaussian proposal density.  The move to the updated
value $q^1$ is accepted with probability min(1,r) where
\[
r = \frac{p(q^1 \vert \{ N_i^1 \})}{p(q^0 \vert \{ N_i^1 \})} .
\]
The procedure is then repeated to obtain successive samples
$\{N_i^2\},q^2$, $\{N_i^3\},q^3$, ....

The Gibbs sampler requires some time to burn-in,
that is, to achieve a stationary distribution.
For this reason some initial number of samples need to be discarded.
One may run a number of Gibbs chains simultaneously, usually from
different starting points, to test convergence and
the coverage of parameter space. 
Successive samples in a Gibbs chain
are somewhat correlated, so it is common to thin the chain by discarding
samples less than a correlation length apart.  The practicalities of
these issues in the current context will be discussed below.

\subsection{Marginalization}

Maps can be made by this procedure from self-calibrated data.
But logically, the complex antenna-based gains, \{$\tilde{g_{\alpha}}$\},
can and should be determined simultaneously along with the map.
These gains are then treated as nuisance parameters
and marginalized in order to obtain the optimum map.
The least informative priors for \{$\tilde{g_{\alpha}}$\}
are uniform in phase
and Jeffreys 1/$\vert \tilde{g_{\alpha}} \vert$ priors in amplitude.
If the original amplitude calibration is believed to be nearly correct,
an amplitude prior distribution centered on
$\vert \tilde{g_{\alpha}} \vert$=1 may be substituted.
But the width of such a distribution should be sufficient
to encompass all realistic gain amplitude variations.
This has the potential to provide a distinct improvement over
current techniques in which the self-calibration step determines
fixed values for the complex gains within some time interval.
If such values are uncertain, treating them as known
can lead to seriously degraded maps.
However, treating them as unknown and uncertain parameters
in the measurement matrix allows one to determine and then
marginalize over their individual probability distributions.
In the Bayesian context,
this is the correct way to deal with such parameters.
Other instrumental parameters may be treated in a similar manner.

\subsection{Advantages}

The above procedure requires no gridding, interpolation,
or averaging of the visibility data,
practices which have been shown by \cite{B95} to cause problems in the
CLEAN procedure.  In Bayesian analysis such practices are obviously
incorrect as they involve a loss or corruption of information.
The procedure also does not require an \emph{ad hoc} choice
of weighting schemes, whether natural or uniform weighting,
Briggs robust weighting \citep{B95}, or otherwise. 
The Bayesian analysis automatically makes optimal use of each
visibility measurement, in the information-theoretic sense, without
requiring the user to make tradeoffs.  These advantages, however, 
require considerably greater computational resources.

Since we avoid gridding and therefore do not use 
the fast Fourier transform (FFT),
the measurement matrix may be set up exactly, including the w component
of the interferometer baseline along the line of sight,
at little additional cost.
Visibilities then are no longer Fourier transforms of the intensity
distribution.  Instead the transform kernel is
\[
\frac{exp\{-i2\pi[\xi u+\eta v+w (\sqrt{1-\xi^2-\eta^2}-1)]\}}{\sqrt{1-\xi^2-\eta^2}} \; .
\]

\section{Numerical Results and Tests}

Given the variety of situations that can arise in radio astronomy,
no single example can prove the general superiority of one technique
of data reduction over another.
We believe that the arguments presented above
show the theoretical advantages of our method.
Here we present an example for the purpose of illustrating
its operation and its practicality.

\subsection{Test Object}

We simulate observations at a frequency of 100 GHz
for a region of the sky containing 8 elliptical Gaussian components
as shown in Table 1 and Figure 1a.
The parameters were drawn from random distributions.
The logarithms of the component fluxes were chosen
from a distribution from 0.05 to 5 Jy.  
The component centers were chosen within a circle of radius 58 arcsec,
representing the half-power beam of the BIMA telescopes at this frequency.
Semi-major axes were chosen from a logarithmic
distribution from 1 to 10 arcsec. 
The resulting peak surface brightnesses varied by over 2.5 orders of magnitude.

\subsection{Model Data}

Artificial $u,v$ data were generated by the Miriad task $uvgen$
to simulate the process of observing with the BIMA B-array at 100 GHz.
The observations were assumed to be taken over an hour
angle range of $-$4 to +4 hours, with 2 minute integrations
separated by 2 minute gaps.
The center of the field of view was taken to be 30 degrees north
declination, and the latitude was 40 degrees north.
Minimum and maximum $u,v$ distances were 4.46 k$\lambda$ and 78.00 k$\lambda$,
as shown by the $u,v$ sampling pattern in Figure 2.
Additive, independent Gaussian noise corresponding to fixed system temperatures
of 120 K was included for each integration.
A single continuum channel with bandwidth of 1000 MHz was taken.
The effective area of the telescope was assumed to correspond to 150 Jy/K.
No systematic, antenna gain or system temperature fluctuations were included.
The resulting set of 5490 complex visibilities was then taken to be the data set
for the observations, to be processed by the Bayesian algorithm.
Figure 1b shows a principal solution map for these observations, using
natural weighting, to illustrate the level of sidelobe contamination present.

\subsection{Pixels, Units, etc.}

The effective resolution of the BIMA B array was of order 2 arcsec.
To satisfy the sampling theorem we require a pixel size of
\[
\Delta \xi = \Delta \eta < \frac{1}{2} \, \frac{\lambda}{B_{max}} \,
rad \, \approx 1.03 \times 10^5 \; \frac{\lambda}{B_{max}} \; arcsec \; ,
\]
where $B_{max}$ is the maximum projected baseline length.
So in this case we choose
\[
\Delta \xi = \Delta \eta = 1 \; arcsec \, .
\]
The pixel solid angle
$\Omega_{pix} = \Delta \xi \Delta \eta = 2.35 \times 10^{-11} \; sr$.
The diameter of the main beam is 115 arcsec.
The support is twice this diameter, yielding a total of 41545 pixels.
The unit of brightness is Jy per steradian. 
One may multiply by $\Omega_{pix}$ to get Jy per pixel.
The brightness is related to the usual radio astronomy
Rayleigh-Jeans brightness temperature by the relation
\[
T_b = \frac{\lambda^2}{2k_b} \, I(\xi,\eta) \, .
\]
All integrals over the brightness distribution
become sums over the pixels.

Ground-based single-dish measurements are usually made
in a differential mapping mode. 
Therefore, any spatially uniform brightness is not measured. 
The same is true of interferometric measurements.
However, as discussed above in section 2.3, the true zero
of intensity should be taken into account.
At millimeter wavelengths such a uniform background brightness distribution is
present in the form of the cosmic microwave background (CMB).
We require that the total intensity be positive definite,
not the intensity relative to the CMB.
Taking $T_{CMB} = 2.726 \, K$, the brightness of the CMB is given by
\[
T_b(\nu) = \frac{h\nu}{k_b} \frac{1}{e^{h\nu / k_b T_{CMB}}-1} \, .
\]
At 100 GHz this corresponds to a brightness temperature
of 1.00 K and an intensity of 7.2 mJy arcsec$^{-2}$.
The presence of the CMB is not currently taken into account.

\subsection{Results}

First the data were reduced using the \emph{clean}
and \emph{maxen} (maximum entropy) tasks in Miriad.
Since these methods have adjustable parameters,
there is no unique answer for either technique. 
But parameters were chosen that seemed appropriate
for these test data (e.g. robust=$-$1 in \emph{invert}
and rms=0.001 Jy/beam in \emph{maxen}).
CLEAN produces both positive and negative artifacts,
whereas MAXEN produces only positive artifacts.
CLEAN has no knowledge of there being an absolute zero of intensity.
MAXEN does, but has no knowledge of uniform background emission from the CMB.

We also used a non-negative least squares (NNLS) technique.
Again, "non-negative" should refer to the true zero of intensity,
although no information about uniform background emission
has been included here.
With 41545 pixels and 5490 complex visibility data
such a map is underconstrained,
but non-negativity is a strong regularizer.
The underconstraint problem could be further ameliorated by removing
pixels beyond the half-power beam radius (a more compact support)
or by using fewer, larger pixels.
Instead, we further regularize the NNLS solution by adding
a multiplicity-like term.  The gamma function may be used as an
analytic replacement for the factorials in our earlier development.
We refer to this technique as Gamma-regularized NNLS ($\Gamma$-NNLS).
Further details on this regularizer are given in Appendix C.
By this means we achieve a continuous (as opposed to discrete)
multi-dimensional space in which the figure-of-merit is differentiable.
This allows us to use a conjugate-gradient method of optimization.
It must be emphasized that this technique itself is not
Bayesian, although it has many close similarities with our Bayesian method.

The resulting maps are shown in Figure 3.
The CLEAN map, made using the \cite{C80} algorithm, is shown in Figure 3a. 
Sources 2, 4, 5, 6, 7, and 8 are clearly present.
The most compact sources (numbers 4 and 8) are spatially enlarged
by the 2.56 x 2.07 arcsec restoring beam.
The most extended source (number 2) is seen, but neither
its size nor its flux are well reproduced by this technique.
Peak residual artifacts (noise and sidelobes) in the CLEAN map
are at the level of +0.0008 and $-$0.0009 Jy/pixel,
near the level of the lowest contour.
Source 1 may be marginally detected
just below the lowest map contour shown.
There is no evidence of source 3.
Since sources away from the center of the map
are attenuated by the primary beam, the CLEAN map has
been reproduced in Figure 3b after division by the primary beam profile.
Note the increased brightnesses of source 2 and 7, which are
near the half power radius of the primary beam.

The MAXEN map is shown in Figure 3c.
The strong compact sources 4 and 8 are well represented in size
and shape and are fairly well represented in flux.
Sources 5 and 6 are also clearly present.
There are traces of sources 2 and 7, although both are
in regions of strong sidelobe confusion (cf. Fig. 1b).
There are no negative artifacts, and the largest
positive artifact is +0.0017 Jy/pixel.
The overall range of artificial structure in the map (positive to negative)
is therefore the same as that of the CLEAN map.
Maximum entropy maps exhibit a bias towards positive intensities.
The source of this bias is
a logarithmic singularity in the derivative of the configurational entropy
at zero brightness, as illustrated in Appendix C.

The $\Gamma$-NNLS map is shown in Figure 3d.
The peak brightness on source 8 exceeds that 
of both the CLEAN and the MAXEN maps. 
Sources 4, 5, and 6 are also seen.
All of these are well represented in size and shape.
The extended sources 2 and 7 are present; their structures are reproduced
better than in the MAXEN map and similar to the CLEAN map.
The faint source 1 is clearly visible,
and there are even suggestions of source 3.
No negative artifacts are present,
and the largest positive artifact is +0.0045 Jy/pixel,
larger than for the overly smoothed MAXEN map in Figure 3c.
However, values this large are seen only well outside of the primary beam,
practically at the edges of the map.
Recovered fluxes are discussed below.

The Bayesian algorithm was applied to the same data.
There were no adjustable parameters.
To improve the convergence speed, the map may be started with
a reasonable first guess.
We have seen best results starting from our $\Gamma$-NNLS map.
The posterior varies with the level of quantization,
which was scanned separately over a range from 0.001 to 0.025 Jy/pixel.
The map was reoptimized at each quantization level.
The dependences of $\chi^2/2$, $ln \, W - \lambda \, ln \, n_{pix}$,
and the posterior on the quantization level q are shown in Figure 4.
Note that $\lambda$ is approximately proportional to $q^{-1}$,
as expected for fixed total flux.
In this case the maximum posterior density occurs at a level
of about 0.017 Jy/pixel, and 447 quanta are required to make the map.
This map, often called a \emph{maximum a posteriori} (MAP)
estimator, is shown in Figure 5.
We will refer to this as the posterior mode map.
Since we cannot exhaustively search parameter space,
we do not know that this mode is the true posterior maximum,
only that it is the largest local maximum found in our search.
It is likely that configuration space contains many
nearly degenerate modes.

The minimum value of $\chi^2/2$ in Figure 4 is 5520,
which is close to the number of measurements
(5490 complex visibilities, or 10,980 real data)
and within the 68\% confidence range
of expected values for chi-squared.
Given that we restrict ourselves to brightness quantized maps
on 1 arcsec pixels with the posterior maximized at fixed quantization,
one might not have expected such good agreement.

The Bayesian posterior mode map shows compact sources 4, 5, 6, and 8.
The extended sources 2 and 7 are also clearly present,
with distributions as smooth as might be expected
for a coarsely quantized map.
Occam's razor discriminates against finer quantization.
But intermediate brightnesses (for example, values between 0 and 1 quanta
in sources 2 and 7) are effectively achieved by assigning quanta
to a particular fraction of the pixels within a certain region.
There is evidence in the mode map for weak source 1 but not source 3.
The mode map most closely resembles the $\Gamma$-NNLS map
from Figure 3d, as expected due to their mathematical similarity.
Qualitatively the mode map does a much better job with these data
than the MAXEN algorithm (Figure 3c) and better than the CLEAN
algorithm (Figure 3a), considering that the smoothness of the
distribution in the CLEAN map is an artifact of the method of cleaning used.

A quantitative comparison of source fluxes derived using
the various techniques is given in Table 2.
Source fluxes for the CLEAN, MAXEN, and $\Gamma$-NNLS techniques
are derived by gaussian fits.
The CLEAN technique generally recovers at least 70\% of
the component fluxes, except for the weakest
(1 and 3) and most extended (2) sources. 
The MAXEN technique does the best job of reproducing sizes
of strong compact sources, but underestimates the fluxes.
The $\Gamma$-NNLS technique outperforms both CLEAN and MAXEN for this data set.
It recovers a larger fraction of the source fluxes than CLEAN,
although fully unconstrained 6 parameter gaussian fits
are unstable for weak sources 1 and 3.
For the Bayesian mode map, no gaussian fitting is required.
All quanta inside or near the half power primary beam radius are
clearly assignable to individual sources, except for 3 quanta
that might belong to either source 2 or source 5.
We empirically assign one of these to source 5
and the remaining two to source 2.
The division of flux between these overlapping components is
not a proper Bayesian question, or at least is a question
which has not been asked here in a Bayesian fashion.
Using Gibbs sampling we assign 90\% confidence
intervals to the source fluxes as described 
in the next section.

There is one difference in technique used for these flux comparisons
which deserves further note.  As mentioned above, fluxes for
the CLEAN, MAXEN, and $\Gamma$-NNLS techniques are derived by gaussian fits.
Fluxes for the Bayesian mode map and the Bayesian samples
are derived in a manner similar to aperture photometry.
An aperture photometry technique would be difficult to implement
in an unbiassed fashion with CLEAN due to the regions of negative
emission surrounding bright sources.  That is, calculated source fluxes
would often \emph{decrease} with increased aperture size.
For MAXEN and $\Gamma$-NNLS all residuals are positive, so calculated source
fluxes would systematically \emph{increase} with increased aperture size.
On the other hand, for Gibbs sampling any source visible in the
mean map may have its statistical flux distribution calculated directly
from the sampling.  Examples are shown in the next section.

\subsection{Sampling of the posterior density}

The details of the Gibbs sampling process are given in Appendix B.
Figure 6 shows one run of the Gibbs sampler, starting from
the "mode" map of Figure 5.  The first few scans increase the
number of quanta in the map, reduce $\chi^2$, and reduce the
posterior.  These changes are expected and primarily affect
pixels beyond the half-power radius of the main beam.
After about 10 scans a nearly stationary distribution is achieved.
This is about the correlation length for the outer pixels,
and these first 10 scans (the "burn-in") are discarded. 
The remaining 200 scans (in this case)
produce of order 20 independent realizations of the outer
parts of the map.  For these high signal-to-noise data
and for values of q this large, the inner
pixels are tightly constrained by $\chi^2$ and are rarely changed
in the sampling process.  In other words, their coherence lengths
are much longer than 200 scans. 
For smaller values of q, the coherence lengths
for all pixels should be reduced.
In the absence of good evidence for mixing of the Gibbs chains,
we assume that multiple runs in which the sampler is given
different starting points provides an adequate statistical
sampling of parameter space.
The independence of the various starting points of the sampler
depends on the random choice of the quantization level
and the probabilistic rounding employed to determine 
the initial number of quanta in each pixel.

In Figure 7 we show a mean map and a standard deviation map
based on 250 independent runs of the Gibbs sampler,
each of length 200 scans after burn-in.
Except for a few pixels in the highest surface brightness sources,
the square-root of the pixel variance is of the
same order as, or larger than, the mean flux for that pixel.
Such behavior might be expected from examination of the mode map
in Figure 5, where the vast majority of the pixels have either
zero or one quantum of flux.
Figure 7, however, presents only part of the available statistical
information, namely the first and second moments of the flux
distributions in each pixel.  It does not include co-variance information,
and there are strong pixel-to-pixel correlations (and anti-correlations).
An example is shown in Figure 8.

Due to these strong correlations, Figure 7 by itself does not present
enough information to calculate uncertainties in the fluxes of
our 8 sources.  However, such information is available from the
full multi-dimensional probability density.  For each scan of each
run of the Gibbs sampler, values for the source fluxes are recorded.
As with the case of the "mode" map, there is little ambiguity
as to which (if any) of the sources a particular quantum belonged,
except for the overlapping source 2 and 5.  Cumulative probability
distributions are shown in Figure 9, and 90\% confidence limits
are listed in Table 2.
In the case of sources 2 and 5, the total flux was attributed 80\%
to source 2 and 20\% to source 5.
As discussed earlier, the allocation of flux to one or
the other source is not a Bayesian question, since our hypothesis
space knows nothing about "gaussians" or indeed even about the
existence of discrete "sources".

The actual input source fluxes fall within
the 90\% confidence limits for 5 of the 8 sources. 
For two other sources, the true flux lies
just above the 90\% confidence interval. 
A tendency to underestimate flux may be due to a cutoff
at low q adopted for practical reasons (Appendix C).
Suggestions of emission in several other regions
within the half-power beam may be seen in Figure 7.
Fluxes have been checked for 3 of these regions (Fig. 9h,i,j)
and are consistent with zero.

\section{Discussion}

\subsection{Concerns and Cautions}

One supposed drawback of methods incorporating maximum entropy
is that they understate flux densities at peaks in the maps.
This indeed appears to be true for MAXEN, for reasons discussed further
in Appendix C.  This also appears to be true of the Bayesian
posterior \emph{mode} map derived here.
But the full Bayesian posterior density distribution should not
make systematic errors.
We show that the Bayesian posterior \emph{mean} map is a much
better reflection of the actual photometry.
But just as neither the mode, mean, nor median
can express the full nature of a general numerical distribution,
no single map can represent the entire information contained in
the sampled multi-dimensional posterior density.

Another alleged concern with maximum entropy techniques
is the variability of the noise and the spatial resolution across the map. 
This is a true feature of both Bayesian and maximum entropy techniques.
Anyone desiring a fixed resolution map may certainly smooth (convolve)
a map to produce coarser -- but nearly constant -- resolution.
But such a procedure entails loss of information. 
It is simply a fact that one can achieve high spatial resolution
only with high signal-to-noise ratio.

Some have advocated the use of "data-adaptive models,"
usually in default image models.
A data-adaptive model is essentially a data-adaptive prior.
As discussed by \cite{J03} and \cite{Sivia96}, this is an incorrect formulation.
One must not use a posterior as a prior for a second iteration
with the same data.
If the "model" is based on information external to the data set
being studied, it could be incorporated into a prior.
But such a prior may be difficult to formulate exactly.
If possible, a preferable course of action would be
to treat this external information as real data
and incorporate it into the likelihood.

Also remember that we incorporate into our "information" term
knowledge about noise statistics, particularly that they are Gaussian.
Although Gaussian noise is least informative, noise
sources are often non-Gaussian.
It may be desirable to incorporate
a modified penalty function for large fluctuations,
as is often done in robust statistics.
Likewise, non-stationary noise would require a more careful treatment.

An optimal imaging algorithm is not a panacea.
In fact, it places stronger demands on the observer's
knowledge of instrumental characteristics: passbands,
telescope pointings, noise levels, beam shapes, etc.
The quality of the resulting images can easily
be limited by uncertain knowledge of the true measurement matrix.
And the design of an observational program must 
ensure that the information content of the data is as
large as possible, for example by using optimal $u-v$ coverage.

\subsection{Mosaic Imaging}

For mosaic imaging, the region of support should be the union of the single
dish pointings.  In the case of heterogeneous arrays, normally one
would preserve the data inside the largest of the relevant combined beams.
For dissimilar antenna pairs, the beam is the geometric mean of the
individual beams.  Otherwise, this technique should easily generalize
to handle mosaic data.  For modest sized mosaics the number of pixels
will grow less rapidly than the number of pointings, due to overlap.

\subsection{Error Analysis}

At lowest order, error analysis consists of assigning uncertainties
(error bars) or confidence intervals to various derived quantities. 
At a greater level of sophistication one is concerned with
the covariance/correlation matrix and with higher moments
(higher than second) of the probability distribution. 
However, to fully understand the uncertainties present
in a multi-dimensional distribution one needs to formulate
the probability density in any region of parameter space. 
This cannot be done analytically.
But the sampling method described here allows one to approximate
this density, to an accuracy limited only by the amount of available
computing power.  One can then address any well-posed question related
to the posterior distribution.

\subsection{Self Calibration}

Self calibration was discussed briefly in section 2.6.
A problem remaining to be solved is the time variation of the complex gains.
A common practice is to adopt a self-calibration interval
(a time scale for averaging) long enough to provide sufficient
signal-to-noise ratio but short enough to allow for correction of
atmospheric phase fluctuations.  However, these considerations should
be formulated into an appropriate prior for the time-dependence of the gains,
rather than being based on an \emph{ad hoc} averaging time.
The choice of a reference antenna for the phase is also typically made
on an \emph{ad hoc} basis.  This should be treated objectively,
as should the choice between phase-only and phase-and-amplitude correction.
If the signal-to-noise ratio is low, some gains will be poorly known.
Treating the complex gains as nuisance parameters is
the correct course of action.  
The problem of atmospheric or ionospheric isoplanicity \citep{Sch84},
which arises in wide-field imaging,
in principle can also be treated as a Bayesian problem.

\subsection{Pixels Redux}

There are other possible ways to specify pixelations and basis functions.
Although both CLEAN and MEM methods are often described as Fourier inversions,
in our development there is no explicit calculation of Fourier
coefficients.  The visibility data are not gridded in $u-v$ space, and no
advantage is taken of the FFT.  The modified Fourier kernel, though, is used
to calculate the interferometric likelihood.
\cite{PS96} have proposed an MEM approach utilizing wavelets and a
multi-scale measure of entropy.  We do not see any clear advantage
to their approach, due to the need to enforce non-negativity. 
We speculate, though, that there may be
advantage to adopting other pixelations.
Minor gain may be possible with a hexagonal grid of pixels.
However, a much greater advantage may be possible 
by using other pixel sizes or with an adaptive grid approach.  
A Bayesian approach would tell us how many pixels the data justify
and where they should be densest.
Small pixels would be used in
regions of rapidly-changing brightness with high signal-to-noise,
and large pixels would be used elsewhere, such as in the region
beyond the half-power beam radius.
In the example given above, a reduction in the number of pixels
by as much as a factor of 100 may be possible.
The posterior would be affected only logarithmically,
but the computation time would be greatly reduced.

In all of the preceding discussion we assume that
one wishes to produce an image.  However, as pointed out by
\cite{C93}, "if a scientific question can be answered without
making an image, it is best to do so."
In such cases the general Bayesian framework should still be used,
but a different hypothesis space would be required. 
The nature of that hypothesis space and the assignment
of prior probabilities would depend on the scientific question
being addressed.  This is why in CMB analysis, images are of
secondary interest.  For gaussian, isotropic models of the CMB,
all cosmologically relevant information is contained in the 
power spectrum multipole coefficients {$C_l$}.

\section{Conclusions}

We introduce a Bayesian approach to the analysis of radio
interferometric data.  This approach incorporates the key
feature of brightness quantization and a multiplicity prior,
differing thereby from conventional maximum entropy techniques.
It automatically incorporates Occam's razor,
limiting the amount of detail in the map to that justified by the data.
Monte Carlo sampling techniques are used to determine
the multi-dimensional posterior density distribution. 
From this we can determine statistical properties on
a pixel-by-pixel basis, such as a mean map, a variance map,
and correlations between individual pixels.
We can readily deal with heterogeneous arrays,
mosaic data, and single-dish data.
This technique also has the ability
to determine and marginalize over a number of
unknown parameters in the measurement matrix.
This should allow observers to incorporate self calibration
into the map making process in a self consistent fashion.
The ability to marginalize over atmospheric or instrumental
phase fluctuations may be one of the most valuable
aspects of this approach.

This current approach to interferometric imaging arises out of our
prior work in a number of different contexts \citep{S85,S95,W03,W04}
seemingly unrelated to radio interferometry.
This exemplifies, therefore, the growing recognition
within the astronomical community of the importance
of sophisticated statistical techniques, particularly Bayesian analysis.
The range of applicability of such techniques is likely
to grow significantly in the near future.

\acknowledgments
This work was supported in part by the
Laboratory for Astronomical Imaging at the University of Illinois, by the National Science Foundation under grants AST 02-28953 and
AST 05-07676, by NASA under subcontract JPL1236748, by the National
Computational Science Alliance under AST300029N and by the Center for
Advanced Studies at the University of Illinois.

\appendix

\section{Implementation of a Grid Search Algorithm}

An appropriate starting point for Bayesian analysis would seem
to be an algorithm which is effective at finding good (high)
local maxima in the posterior density.  This is more difficult
than it sounds, due to the discreteness of our configuration space.
Finding the global maximum is generally not practical. 
But we do not need the global maximum.
Instead we need a computationally efficient way to find locations
in parameter space with high probability which may be used as
starting points for the Gibbs sampler.  The implementation of the
Gibbs sampler itself is given in Appendix B.
The algorithm described here was used to produce the results shown
in Figures 4 and 5.  Figure 4 was used to estimate optimal values
of the quantization level q and as a way to
generate starting points of the individual Gibbs samplers.

Figure 10 illustrates some potential moves in configuration space.
We describe a move as \emph{axial} in configuration space
if it is a move of unit length along one axis.  In other words,
it involves a single pixel to which a single quantum is added
or, if the pixel is occupied, from which a single quantum is subtracted.
A move is \emph{diagonal} in configuration space
if single quantum moves are made simultaneously along two configuration axes.
In the most common case, such a move is
\emph{flux preserving} and also links adjacent pixels.
In other words, a quantum is moved from one pixel
to one of the eight adjacent pixels in the map.
Other moves are more complicated and we use a shorthand
in which \emph{u} (up) stands for the addition of a quantum
and \emph{d} (down) stands for the subtraction of a quantum.
Our set of additional moves contains \emph{dd}, \emph{dud},
\emph{duu}, \emph{ddd}, \emph{dudu}, and \emph{dududu}. 
The moves \emph{axial} and \emph{dd} have no spatial restrictions.
For computational practicality, the remaining moves are spatially restricted,
generally to sets of pixels with a maximum pairwise separation of three pixels.
For fine quantization (small q), the simpler moves are generally sufficient.
But for coarser quantization, local maxima may be separated by
significant distances in configuration space and the more elaborate
move set is essential.

Calculation of \emph{diagonal} moves is rapid,
due to the limited number of quanta ($\lambda$) present.
So searchs for \emph{diagonal} moves which make
the greatest improvement in the posterior are done frequently.
Next most efficient is the search for \emph{axial} moves
which make the greatest improvement in the posterior.
In configuration space, an \emph{axial} move is a move
to a nearest-neighbor on the lattice.  A \emph{diagonal}
move is a move to a second-nearest neighbor.
In a space of large dimensionality, the volume to be searched
increases rapidly with the length of the move.
Therefore we conduct only very limited searchs of
third-nearest neighbors and higher.

Other techniques may be more efficient. 
In particular, problems of this sort are often well suited
to treatment by "simulated annealing."
For coarse quantization the potential barriers for changing
the innermost pixels are large, requiring large pseudo-temperatures
to "melt" any initial configuration.  
Simulated annealing may be an effective technique to explore
the vast configuration space available at large $\lambda$ (small q).
However, so far we have not found an efficient implementation
for small $\lambda$ (large q).
As discussed earlier, finding the
true maximum of the posterior is not essential.
Since a large fraction of the probability lies at moderate $\lambda$,
a simulated annealing approach may ultimately be most efficient.

\section{Implementation of Gibbs Sampler}

A general outline of the Gibbs sampler was given in section 2.5.
The first step is to choose an initial value for q and an
initial quantized map.  Figure 4 shows that the most
probable maps are found with q equal to about 0.017 Jy/pixel.
A naive approach would be to pick q randomly from a distribution
centered on this posterior maximum and with a width
determined by the curvature of the log posterior.
However, there is a vastly larger volume of configuration space
available at larger $\lambda$ (smaller q).  
The same factor that makes any particular configuration
unlikely at small q, simultaneously says that there will be many more
such configurations.  To provide an unbiassed sampling of the
posterior density, we need to start a significant fraction of
our Gibbs samplers in this "unlikely" region of parameter space.
If the samplers have sufficient mobility, they will automatically
cover parameter space in the correct probabilistic sense.
However, we have some reason to doubt the mobility of the samplers
due to the discreteness (the quantized nature) of the space (see section 3.5)
and the high barriers present at large q.
Mobility is somewhat better at small q.
For now we adopt a truncated gaussian distribution in q
with a mean of 0.009 Jy/pixel and 
a standard deviation of 0.004 Jy/pixel.
The truncation rejects values below 0.003 Jy/pixel,
a region which is computationally inefficient and also less probable.
We do not claim this procedure to be optimal.

Next we need an initial distribution of quanta.  We start from our
$\Gamma$-NNLS map and divide the brightness in each pixel by q.
Then we randomly round fractional
pixel values n.x up to n+1 quanta with probability $0.x$
and down to n quanta with probability $1 - 0.x$.
The search algorithm described in Appendix A is then run
until it encounters a local maximum.

From there the Gibbs sampler is started.  The conditional probabilities
during scan j to change pixel i from its current value $N_i^{j-1}$ to
various possible values $N_i^{j}$ are proportional to
\[
exp [ \Delta \; posterior ] = 
\]
\[
exp [ -\Delta \chi^2 + \Gamma (\lambda+1+\Delta\lambda) - \Gamma (\lambda+1)
    + \Gamma (N_i+1) - \Gamma (N_i+1+\Delta \lambda) 
    - \Delta\lambda \, ln \, n ]
\]
where n is the number of pixels and $\Delta\lambda = \Delta N_i$.
At each step we accumulate the first and second moments of
the probability density of $\lambda N_i$ based on these \emph{conditional}
probability densities.
These probabilities are then normalized to add up to unity,
and a random draw from [0,1] is then used to select $N_i^{j}$
as that value where the sum of the conditional probabilities
exceeds the random number.
At the end of each scan we record the values of each pixel and
the fluxes within any regions of interest.

At the end of each scan the value of the continuous variable q is also updated.
The curvature of the posterior density with respect to q may be
calculated, suggesting a proposal density of width
\[
q \left( \onehalf \sum_{k=1}^{N_{vis}}\frac{ | V_k^{\prime} | ^2}{\sigma_k^2} \right) ^{-1/2} \; .
\]
To encourage q to sample a larger region of parameter space we
double this width. 
From this a new trial q is chosen.
The remaining details of the sampler are described in section 2.5 and 3.5.

There are undoubtedly more efficient and better mixed
types of samplers.  The Gibbs sampler has the disadvantage
of examining one pixel at a time (equivalent to the \emph{axial}
moves described in Appendix A).  As we have seen,
there can be large potential barriers to the addition or
subtraction of individual quanta.  Much better would be
a Markov chain which implemented \emph{diagonal} moves.
One could also envision a sampler which permitted widespread
redistributions of existing quanta, as well as additions
and subtractions of quanta.  
So-called genetic algorithms are often recommended for
discrete parameter spaces, such as those encountered here.
And steps from different procedures may be combined
to make a more efficient sampler.
Additional programming and testing will be
needed to find the best of these possible methods.

\section{Implementation of the $\Gamma$-NNLS Regularizer}

The Gamma function forms the basis of a nearly ideal regularizer
for this problem since it is continuous, differentiable,
and is exactly equal to the factorials at appropriate integer arguments.
However, it has two notable deficiencies.
Firstly, $\Gamma(x+1)$ is finite for negative x as long as $x > -1$.
Such arguments x correspond to negative brightnesses and are unphysical.
Secondly, $ln \; \Gamma(x+1)$ has a local minimum at x = 0.46163,
which does not correspond to any feature of the physical problem.
However, these problems can be solved by some simple modifications.

For our regularizer R(x) above x = 2 we will use the 
logarithm of the Gamma function,
\[
R(x) = ln \; \Gamma (x+1) \;\;\;\;\;\; x \geq 2 \; .
\]
Between one and two we will add the term shown below,
\begin{eqnarray*}
R(x) & = & ln \; \Gamma (x+1) - (\gamma-1) \, (x-1) (x-2)^3 \\
     & = & ln \; \Gamma (x+1) - (\gamma-1) \, (x^4 - 7x^3 + 18x^2 - 20x + 8) \; \;
\;\;\;\;\; 1 \leq x \leq 2 \; ,
\end{eqnarray*}
where $\gamma$ is Euler's constant.
Between zero and one the regularizer is zero.
\[
R(x) = 0 \;\;\;\;\;\; 0 \leq x \leq 1 \; .
\]
Negative values of x encountered in the optimization process
are handled by the method of Projection onto Convex Sets \citep{B90},
by setting x = 0.
This regularizer, shown in Figure 11,
is continuous, differentiable, and equal to ln x! for
non-negative integers, x.  It has the additional advantage
of being a convex function of x.

The $\Gamma$-NNLS technique is implemented by adopting a
penalty function of
\[
\frac{\chi^2}{2} + \frac{F}{q}\,ln\,n_{pix} - ln \, \Gamma (\frac{F}{q}+1)
+ \sum_i R(\frac{f_i}{q})
\]
by direct analogy with the Bayesian method, with F being the total flux,
$f_i$ the flux in the $i^{th}$ pixel, and q being the ratio between
the flux scale and the arguments of the Gamma functions (thereby
playing the same role as the "quantization" level q, although
here we allow continuous values of the flux in each pixel).
The ratio F/q is analogous to $\lambda$, the number of
quanta in the Bayesian method.  
The value of q should be chosen, as in maximum entropy treatments,
to give a reasonable value of $\chi^2$.
For this problem we chose q = 0.005 Jy/pixel,
although the result is not very sensitive to this choice of q.
The penalty function is minimized by varying the {$f_i$}.
The terms which incorporated Occam's razor in the Bayesian technique
here becomes a restriction on the total flux F, being dominated
(for $F \la q\,n_{pix}$) by the term $F\,q^{-1}\,ln\,n_{pix}$,
which penalizes additional flux.
This technique converges to the same solution whether starting
from a CLEAN component map or a uniform map.

Note the qualitative and quantitative differences between the Gamma
function and the entropy function $x \; ln x$ in Figure 11.
The logarithm of the leading term in Stirling's formula for x!
is $ \onehalf ln(2\pi x) + x\,ln\,x -x$, which
asymptotically approaches $\Gamma (x+1)$ for large x. 
But at x=1 it gives a value of $ln (\sqrt{2\pi}/e) = -0.081$ instead of 0.
The term $-x$ cancels in the calculation of the multiplicity
and may be ignored, since $ \lambda = \sum N_i \;$.
Further ignoring the term $ \onehalf ln(2\pi x)$ removes
the singularity at x=0, and leaves the usual form of the 
entropy function, x ln x.
The derivative of this entropy function
is logarithmically divergent at x = 0.  
Therefore maximum entropy techniques will not
allow \emph{any} pixel to remain at zero brightness. 
The entropy function has a minimum
of depth 1/e at an argument of x = 1/e.  This is several times deeper
than the dip in the Gamma function, and, as noted above, even
the dip in the Gamma function has no basis in the physical problem.
Unless otherwise constrained by the data, a pixel in a maximum entropy
map will seek out a brightness level of $q/e = F/(e\lambda )$.
The entropy function is narrower and deeper than
the Gamma function, indicating that maximum entropy will generally
produce overly smoothed maps.

{}

\clearpage

\input{tab1.tex}

\input{tab2.tex}

\clearpage

\plotone{f1.eps}
\figcaption{
a) Test object consisting of 8 elliptical gaussian components.
Contour levels are .0005, .001, .002, .005, .01, .02, .05, .1, .2 Jy/pixel.
b) Principal solution map with natural weighting.
Greatest positive and negative sidelobes of the synthesized beam
are +15\% and $-$7\%.
Largest artifacts in the map are +20\% and $-$12\% of peak brightness.
Contour levels are multiples of .035 Jy/beam (approximately .004 Jy/pixel).
Only sources 4 and 8 are clearly seen in the principal solution (dirty) map.
}

\plotone{f2.eps}
\figcaption{
Sampling of the Fourier plane in the simulation.
}

\plotone{f3.eps}
\figcaption{
a) Map produced by CLEAN algorithm in MIRIAD,
restored with a 2.56 x 2.07 arcsec beam.
b) Same map produced by CLEAN algorithm, divided by the primary beam
in order to correct the photometry.
c) Map produced by MAXEN algorithm in MIRIAD.
d) The $\Gamma$-NNLS map discussed in the text.
In all plots the contour levels are .001, .002, .005, .01, .02,
.05, .1, and .2 Jy/pixel.  Negative contours at the same levels
are provided for the CLEAN maps.
}

\plotone{f4.eps}
\figcaption{
a) Variation of the posterior density and the terms $-\chi^2/2$
and $ln \, W - \lambda \, ln \, n$ with the quantization level q.
b) Optimal number of brightness quanta as a function of the
quantization level.
}

\plotone{f5.eps}
\figcaption{
Posterior mode map for a quantization level of 0.017 Jy/pixel.
}

\plotone{f6.eps}
\figcaption{
Sample run of Gibbs sampler.
}

\plotone{f7.eps}
\figcaption{
Maps based on moments of the pixel by pixel
conditional probability density distributions
from 250 independent Gibbs samplers of 200 steps each.
The mean map is shown in color and contour levels in parts a and c.
The standard deviation map in parts b and d uses the same colors
and contours as the mean map.
The contour levels are .001, .002, .005, .01, .02,
.05, .1, and .2 Jy/pixel, the same as for Figure 3.
The color scale runs from blue to red (0.005 Jy/pixel)
and then saturates on white (0.01 Jy/pixel).
}

\plotone{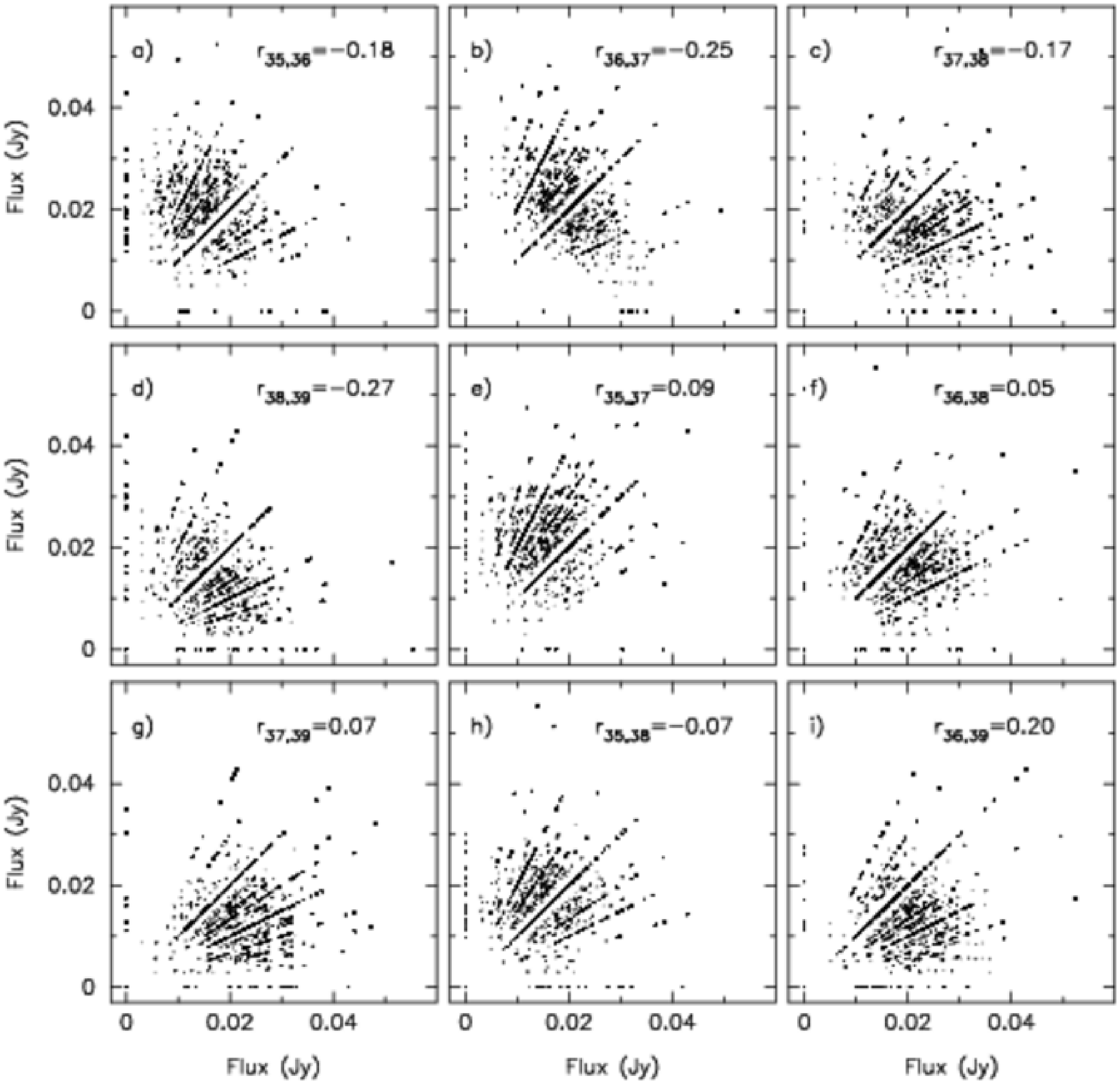}
\figcaption{
Linear correlation coefficients (Pearson's r) for pixel pairs on
a strip of 5 pixels running across
the center of source 5 ($\Delta\delta = 21, \Delta\alpha = 35, 36, 37, 38, 39$).
Panels a, b, c, and d show anti-correlations of adjacent pixels.
The magnitudes of the correlations are reduced
for spacings of 2 arcsec (panels e, f, \& g)
and 3 arcsec (panels h \& i).
The diagonal lines in the plots are artifacts of the quantization
and have slopes of small integer ratios.
}

\plotone{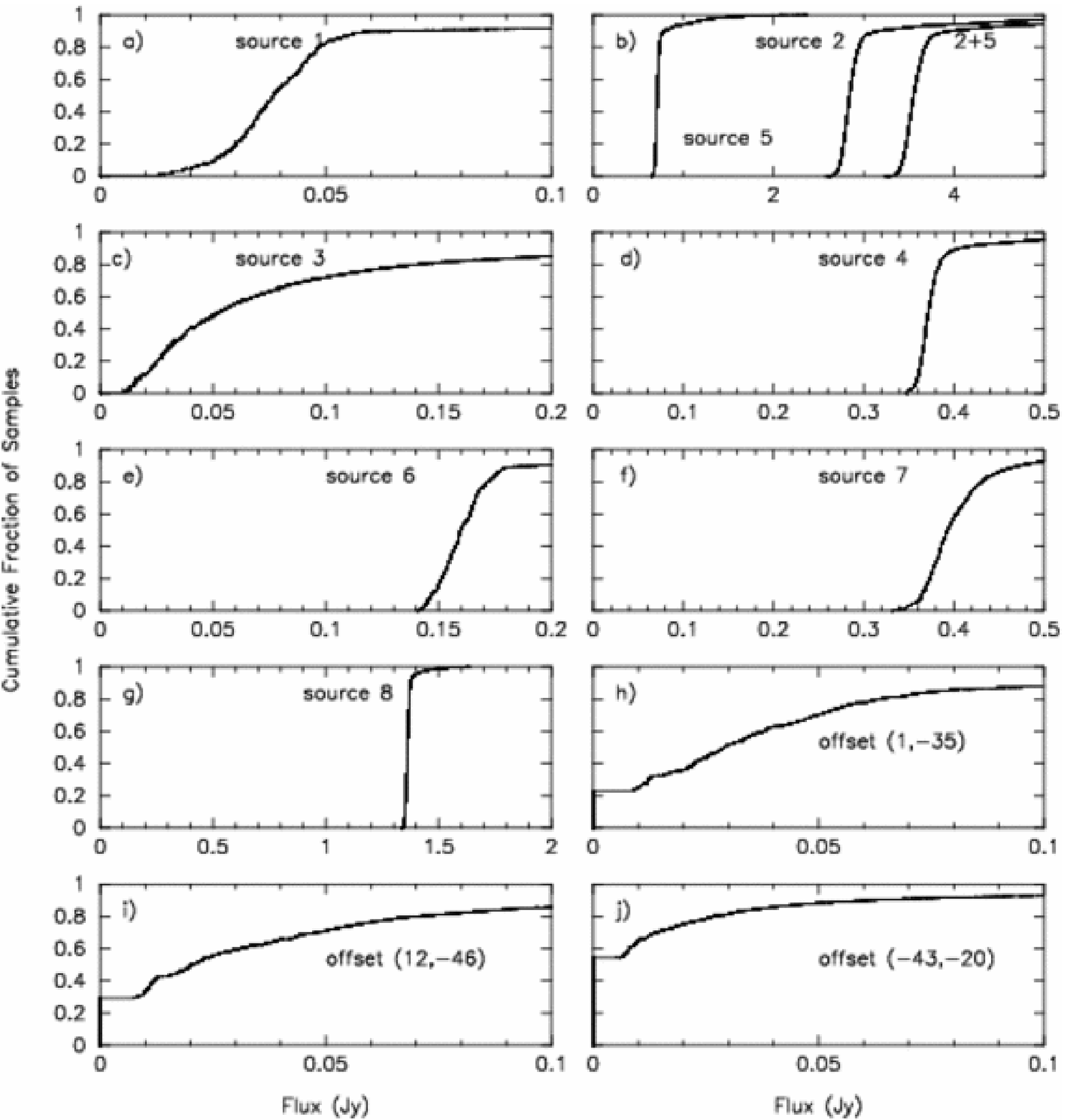}
\figcaption{
Cumulative flux distributions for the eight input sources. 
Also shown are cumulative distributions
for three phantom sources weakly present in Figure 7
at the right ascension and declination offsets shown.
These latter distributions are all consistent with zero at
the 90\% confidence level.
}
\clearpage

\plotone{f10.eps}
\figcaption{
Configuration space for representative pixels i, j, and k.
Points represent allowed (quantized) values on the map.
Circles indicate the plane k=0 and squares the plane k=1.
Representative moves are shown: "axial" moves of unit length
(solid line); "diagonal" moves of length $\sqrt{2}$ (dashed lines)
of both flux conserving (left) and non-flux conserving (right) types;
and, as an example of a move coupling 3 or more pixels, a "body diagonal"
of length $\sqrt{3}$ (dotted line).
}

\plotone{f11.eps}
\figcaption{
Comparison of the regularizer R(x) (solid bold line) discussed in Appendix C
with the logarithm of the Gamma function $\Gamma$(x+1) (dot-dashed line).
The solid points represent ln x! for integer values of x.
Both functions give the correct result at these points.
The region $x < 0$ is unphysical.
The dotted lines represent 2 related functions:
a) $\onehalf ln (2 \pi x) + x ln x - x$,
the logarithm of the leading term in Stirling's formula for x!;
b) $x \, ln \, x$, the usual form of the configurational entropy
(note the minimum at x = $e^{-1}$ and the logarithmically
divergent derivative at x = 0).
}

\end{document}

%% file: tab1.tex
\begin{deluxetable}{ccrrccr}
\tabletypesize{\small}
\tablewidth{0pt}
\tablecolumns{7}
\tablenum{1}
\tablecaption{Test Object Parameters}
\tablehead{
\colhead{Source} & \colhead{Flux} & \colhead{$\Delta \alpha$} & \colhead{$\Delta \delta$} & \colhead{a} & \colhead{b} & \colhead{pa} \\
 & \colhead{(Jy)} & \colhead{(arcsec)} & \colhead{(arcsec)} & \colhead{(arcsec)} & \colhead{(arcsec)} & \colhead{(degrees)} }
\startdata

1 &0.105 &$-$20.42 &   19.09 &4.09 &3.14 &   38.07\\
2 &3.772 &   40.98 &   35.96 &9.37 &8.51 &   36.67\\
3 &0.093 &$-$43.87 &   10.15 &7.36 &3.97 &   69.62\\
4 &0.379 & $-$8.11 &$-$48.52 &1.21 &0.77 &   27.07\\
5 &0.718 &   37.16 &   21.20 &2.76 &2.61 &$-$77.24\\
6 &0.177 &$-$13.66 &$-$13.47 &2.32 &1.35 &$-$69.62\\
7 &0.419 &$-$38.29 &$-$39.88 &3.44 &3.30 &$-$60.07\\
8 &1.393 &   24.17 &$-$16.21 &1.43 &0.79 &$-$56.79\\

\enddata

\end{deluxetable}

%% file: tab2.tex
\begin{deluxetable}{cccccccc}
\tabletypesize{\small}
\tablewidth{0pt}
\tablecolumns{7}
\tablenum{2}
\tablecaption{Recovered Flux Comparison}
\tablehead{
\colhead{Source} & \colhead{Actual Flux} & \colhead{CLEAN} & \colhead{CLEAN\tablenotemark{a}} & \colhead{MAXEN} & \colhead{$\Gamma$-NNLS} & \colhead{Bayesian\tablenotemark{b}} & \colhead{Bayesian\tablenotemark{c}} \\
 & \colhead{(Jy)} & \colhead{(Jy)} & \colhead{(Jy)} & \colhead{(Jy)} & \colhead{(Jy)} & \colhead{(Jy)} & \colhead{(Jy)} }
\startdata

1 & 0.105 & 0.024   & 0.029   & \nodata   & 0.101\tablenotemark{d}, 0.044\tablenotemark{e} & 0.017 & 0.013--0.072\\
2 & 3.772 & 0.541   & 1.001   & 0.356   & 2.361 & 2.203\tablenotemark{f} & 2.646--3.189\tablenotemark{f}\\
3 & 0.093 & \nodata & \nodata & \nodata & 0.066\tablenotemark{g} & \nodata     & 0.011--0.486\\
4 & 0.379 & 0.198   & 0.330   & 0.166   & 0.391 & 0.376 & 0.348--0.406\\
5 & 0.718 & 0.347   & 0.520   & 0.165   & 0.666 & 0.700\tablenotemark{f} & 0.662--0.797\tablenotemark{f}\\
6 & 0.177 & 0.117   & 0.126   & 0.068   & 0.178 & 0.154 & 0.140--0.193\\
7 & 0.419 & 0.170   & 0.320   & 0.133   & 0.405 & 0.376 & 0.332--0.473\\
8 & 1.393 & 1.111   & 1.328   & 1.017   & 1.418 & 1.366 & 1.347--1.378\\

\enddata
\tablenotetext{a}{Corrected for primary beam attenuation}
\tablenotetext{b}{Mode map}
\tablenotetext{c}{90\% confidence limits in regions of
greatest posterior density (i.e. smallest intervals containing 90\%
of the posterior density).  Since this is a Bayesian calculation,
these are really \emph{credible} intervals,
but astronomers are probably more accustomed
to the term \emph{confidence} interval.}
\tablenotetext{d}{Gaussian fit stabilized by fixing major and minor axis lengths.}
\tablenotetext{e}{Flux within 5 arcsec radius}
\tablenotetext{f}{Empirical division of flux between overlapping components}
\tablenotetext{g}{Flux within 10 arcsec radius}

\end{deluxetable}